\documentclass[useAMS,usenatbibi,usegraphicx]{mn2e}
\usepackage{psfig}

\newcommand\beq{\begin{equation}}
\newcommand\eeq{\end{equation}}
\def\msun{\,{\rm M_\odot}}
\def\etal{{et al.\ }}
\def\gsim{ \lower .75ex \hbox{$\sim$} \llap{\raise .27ex \hbox{$>$}} }
\def\lsim{ \lower .75ex\hbox{$\sim$} \llap{\raise .27ex \hbox{$<$}} }

\title[Gravitational waves from massive black hole binaries]
{Extreme recoils: impact on the detection of gravitational waves from massive
  black hole binaries.}

\author[A. Sesana]{Alberto Sesana\\
School of Physics and Astornomy, University of Birmingham, 
Edgbaston, Birmingham, B15 2TT, UK}

\begin{document}

\date{Received ---}

\maketitle

\begin{abstract}
Recent numerical simulations of coalescences of highly spinning massive black hole 
binaries (MBHBs) suggest that the remnant can suffer a recoil velocity 
of the order of few thousands km/s. 
We study here, by means of dedicated 
simulations of black holes build--up, how such extreme recoils 
could affect the cosmological coalescence rate of MBHBs, 
placing a robust {\it lower limit} for
the predicted number of gravitational wave (GW) sources
detectable  by future space--borne missions 
(such as {\it LISA}). We consider two main 
routes for black hole formation: one where seeds are 
light remnants of Population III stars ($\simeq 10^2 \msun$),  
and one where seeds are much heavier ($\gsim 10^4 \msun$), 
formed via the direct gas collapse in primordial nuclear disks. 
We find that extreme recoil velocities do not compromise
the efficient MBHB detection by {\it LISA}. If seeds are already 
massive and/or relatively
rare, the detection rate is reduced by only $\sim 15\%$. 
The number of detections drops substantially
(by $\sim60\%$) if seeds are instead light and abundant, but in this case
the number of predicted coalescences is so high that at least 
$\sim10$ sources in a three year observation are guaranteed.
\end{abstract}


\section{Introduction}
Massive black hole (MBH) binaries (MBHBs) are among the primary candidate 
sources of gravitational waves (GWs) at mHz frequencies,
the range probed by the space-based {\it Laser Interferometer Space Antenna} 
({\it LISA}, Bender \etal 1998). Today, MBHs are ubiquitous in the 
nuclei of nearby galaxies (see, e.g., Magorrian et al. 1998). 
If MBHs were also common in the past, and if their host 
galaxies experienced multiple mergers during their lifetime, as dictated by 
popular cold dark matter hierarchical cosmologies, then MBHBs inevitably 
formed  in large numbers during cosmic history. 
Provided MBHBs do not ``stall'', 
their inspiral driven by radiation reaction 
follows the merger of galaxies and protogalactic structures at high redshifts. 
MBHBs coalescing in less than a Hubble time would give origin to the 
loudest GW signals in the Universe, and
a low--frequency detector like {\it LISA} will be sensitive to 
GWs from binaries with total masses in the range 
$10^3-10^7\,\msun$ out to $z\approx 20$ (Hughes 2002).

The formation and evolution of MBHs has been investigated recently by several 
groups in the framework of hierarchical clustering cosmology
(e.g. Menou, Haiman \& Narayanan 2001, Volonteri Haardt \& Madau 2003,
Koushiappas, Bullock \& Dekel 2004). The inferred 
{\it LISA} detection rate, ranging 
from a few to a few hundred per 
year, were derived in a number of papers (Jaffe \& Backer 2003,
Wyithe \& Loeb 2003, Sesana \etal 2004, Sesana \etal 2005, 
Enoki et al. 2004, Rhook \& Wyithe 2005). 
More recently Sesana Volonteri \& Haardt (2007) investigated 
the imprint of massive black hole formation models on the
expected MBHB coalescence rate, finding that at least $\sim 10$
(considering a model that marginally reproduces the observational constrains
and that can be taken as a robust lower bound) sources should
be safely regarded as observable by {\it LISA}, assuming a 3 year lifetime mission.

GWs emitted during the final
plunge of the binary, carry away a net linear momentum, causing a 
recoil of the MBHB center of mass in the opposite direction (Redmount 
\& Rees 1989). This GW recoil could have interesting astrophysical
effects, since many coalescence remnants can be ejected from their
host galaxies and dark matter halos (e.g. Madau et al. 2004, Merritt et
al. 2004, Micic Abel \& Sigurdsson 2006). 
This justifies the increasing effort to 
obtain accurate estimates of the recoil velocity. In the case of 
non spinning black holes, the latest analytical 
(e.g. Favata Hughes \& Holz 2004,
Blanchet Qusailah \& Will 2005, Damour \&
Gopakumar 2006) and numerical (Baker et al. 2006, 
Gonzalez et al. 2007) approaches are now both converging 
to maximum recoil velocities $v_r$ in the range 100-250 km/s
for binaries with mass ratio $q=m_2/m_1\sim 0.4$ ($m_2<m_1$ are the masses of
the two binary members). The expected values are only slightly
higher if the binary is eccentric; Sopuerta Yunes and Laguna (2007)  
found $v_r\propto (1+e)$. 

On the other hand, recent relativistic numerical simulations of spinning 
black hole binaries (Herrmann et al. 2007, Schnittman \& Buonanno 2007) 
suggest that $v_r$ increases linearly with the black hole spin parameter $a$ , 
where $0\le a \le 1$, 
and in the case of highly spinning black holes ($a > 0.8$)
the magnitude of the kick suffered by the remnant could be 
of the order of a few thousand km/s (Tichy \&
Marronetti 2007). Campanelli et al. (2007) report values of $v_r$ as high as 
$\sim 4000$ km/s for equal mass binaries, if both spins lie in the 
binary orbital plane. Such a kick is sufficient to eject the 
remnant not only from a dwarf galaxy, 
where the escape velocity is $\sim 300$ km/s, but
even from the center of a giant elliptical, for which the
escape velocity can reach $2000$ km/s.

Though it is likely that MBHs acquire high spins
(e.g. Volonteri et al. 2005) during their accretion history, 
the impact of the resulting recoil on the MBH assembly has never been 
studied in details so far. 
If extreme recoil is indeed the rule, the ejection 
of a large fraction of MBHs formed through the coalescence of a binary systems 
can cause a significant drop in the number of expected coalescing events 
on the way of MBH assembly. Volonteri (2007) recently showed that current 
assembly models are able to reproduce the major observational 
constraints even if the extreme recoil prescription by
Campanelli et al. (2007) is taken into account. 
However, high kick velocities could seriously affect the expected
number counts predicted for {\it LISA}, since the ejection of 
remnants by their host halos would avoid subsequent MBHB formation.

In this letter we estimate a robust {\it lower limit} for
the predicted number of {\it LISA} sources. We use the Montecarlo 
realizations of the merger history performed by Volonteri (2007) 
to show that even in the worse 
(for GW observations) case scenario in which {\it during each merger} 
the two MBH spins are counter-aligned in the MBHB orbital plane and 
extreme recoil is at work, current MBH assembly models predict 
that at least ten sources will be detectable by {\it LISA}.
In practice, the lower limit of 
the expected number of {\it LISA} sources does not 
substantially drop with respect to models employing non-spinning
MBH recoil prescriptions (e.g. Volonteri Haardt \& Madau 2003). 

\section{Models of black hole formation}
In the hierarchical assembly framework, MBHs form growing 
trough mergers and accretion from seed black holes at high redshift.
There are two main scenarios for MBH assembly,
namely the light seed and the heavy seed models.
In the light seed models, seed MBHs typically form with masses 
$m_{\rm seed}\sim$ few$\times10^2\msun$, in 
halos collapsing at $z\sim20$, and are thought to be the end--product 
of the first generation of stars (Madau \& Rees 2001). 
In the heavy seed models, black hole seeds form already massive  
($10^4-10^5\msun$) from the low angular momentum tail of 
gas in protogalaxies at high redshifts. The angular
momentum distribution of the gas in early-forming halos
can be determined by means of cosmological {\it N}-body
simulations (Bullock et al. 2001); halos with low spin parameters 
are prone to global dynamical instabilities, leading to the 
formation of a massive seed black hole (Koushiappas et al. 2004,
Begelman Volonteri \& Rees 2006, Lodato \& Natarajan 2006). 

We focus here on the two specific models discussed in Volonteri (2007)
that are representative of these two classes of MBH assembly scenarios: 
the VHM and the BVRlf models. 
In the VHM model, representative of the light seed scenarios, 
(Sesana et al. 2007 for details) seed MBHs form with masses $m_{\rm seed}\sim$ 
few$\times10^2\msun$, in halos collapsing at $z=20$ from rare 
3.5-$\sigma$ peaks of the primordial density field. 
In the BVRlf model, representative of the heavy seed scenarios,
(Sesana et al. 2007 for details),  black hole seeds form in halos  
subject to runaway gravitational instabilities, via the so-called 
``bars within bars" mechanism (Shlosman, Frank \& Begelman 1989).
MBH seed formation is assumed to be efficient only in metal free 
halos with virial temperatures 
$T_{\rm vir} \gsim 10^4$K, leading to a population of massive seed 
black holes with $m_{\rm seed}\sim$ few$\times10^4\msun$.

The subsequent MBH evolution relies only on a few simple
assumptions. Nuclear activity is triggered by halo mergers: 
in each major merger the more massive hole accretes gas until its mass scales 
with the fifth power of the circular velocity of the host halo,  
normalized to reproduce the observed local correlation 
between MBH mass and velocity dispersion ($m_{\rm BH}-\sigma_*$ relation).
MBHB coalescence is assumed to occur efficiently following halo mergers. 

For both the VHM and the BVRlf models, we consider two cases that bound the
effect of recoil in the assembly of MBHs and, as a consequence, LISA events: 
(i) no gravitational recoil takes place and (ii) maximal gravitational recoil 
is associated to every MBHB merger, using the model by Volonteri (2007), which 
is based on the estimates reported by Campanelli at al. (2007). For the latter
we use the merger tree realizations presented in Volonteri 
(2007). The model takes into account consistently for the cosmic evolution
of the mass ratio distribution of merging binaries and 
of their spin parameters (see discussion in Volonteri 2007). 
%The spin parameters are 
%determined by the sequence of coalescences and accretion episodes
%experienced by each MBH during its assembly .
In each single merger, the mass ratio and the MBH spin magnitudes
are therefore fixed by the merger hierarchy; 
the spin orientations are instead chosen so as to maximize the recoil.
MBH spins are assumed to initially 
lie in the binary orbital plane, counter-aligned one to each other.
The recoil velocity is then evaluated according to equation 1 of 
Campanelli et al. (2007), that in this case 
simplifies as: 
\begin{eqnarray}
\vec{v}_r(q,a_i)&=& A\frac{q^2(1-q)}{(1+q)^5} \left[1+B\frac{q}{(1+q)^2}\right]
\hat{e}_\|\nonumber\\ &+&K{\rm cos}(\Theta-\Theta_0)\frac{q^2}{(1+q)^5}(a_2+qa_1)\,\hat{e}_\bot.
\end{eqnarray}
Here $A=1.2\times10^4$ km/s, $B=-0.93$, $K=6\times10^4$ km/s, $a_1$ and $a_2$
are the magnitudes of the spin parameters of the two holes, $\hat{e}_\|$ is
a unit vector in the binary orbital plane and $\hat{e}_\bot$ defines 
the direction perpendicular to the orbital plane. 
The component of $\vec{v}_r$ along the $\hat{e}_\bot$ direction 
depends sinusoidally upon the angle $\Theta$ between the MBH spins 
and their initial linear momenta. To get the maximum recoil we set 
$\Theta=\Theta_0$ ($\simeq 0.184$).

We would like to emphasize that the prescription that we have chosen for (ii), 
and whose main features we have just summarized is the least favourable 
for gravitational wave observations and (probably) unlikely to occur in 
these extreme circumstances during MBH assembly 
(Bogdanovic Reynolds \& Miller 2007). 

\section{Gravitational wave signal}
Full discussion of the GW signal produced by an inspiraling 
MBHB can be found in Sesana \etal 2005, along with all the relevant references.
Here we just recall that a MBH binary at (comoving) distance $r(z)$
with chirp mass ${\cal M}=m_1^{3/5}m_2^{3/5}/(m_1+m_2)^{1/5}$ generates
a GW signal with a characteristic strain given by (Sesana et al. 2005):
%%%%%%%%%%%%%%%%%%%%%%%%%%%%%%%
\begin{equation}
h_c=\frac{1}{3^{1/2}\pi^{2/3}}\,\frac{G^{5/6}
{{\cal M}}^{5/6}}{c^{3/2} r(z)}\,f_r^{-1/6}.
\label{eq1h_c}
\end{equation}
%%%%%%%%%%%%%%%%%%%%%%%%%%%%%%%
An inspiraling binary is then detected if the signal-to-noise ratio ($S/N$)
{\it integrated over the observation} is larger than a given 
detection threshold, 
where the optimal $S/N$ is given by
(Flanagan \& Hughes 1998)
%%%%%%%%%%%%%%%%%%%%%%%%%%%%%%%
\begin{equation}
S/N=\sqrt{ \int d\ln f' \, \left[
\frac{h_c(f'_r)}{h_{\rm rms}(f')} \right]^2}. 
\label{eqSN}
\end{equation}
%%%%%%%%%%%%%%%%%%%%%%%%%%%%%%%
Here, $f=f_r/(1+z)$ is the (observed) frequency emitted at time 
$t=0$ of the observation, and the integral is performed over the
frequency interval spanned by the shifting binary
during the observational time. 
Finally, $h_{\rm rms}=\sqrt{5fS_h(f)}$ is the effective rms noise of the 
instrument; $S_h(f)$ is the one-sided noise spectral density, and the
factor $\sqrt{5}$ takes into account for the random directions and 
orientation of the wave; $h_{\rm rms}$ is 
obtained by adding in the instrumental noise contribution 
(given by e.g. the Larson's online sensitivity curve generator 
http://www.srl.caltech.edu/$\sim$shane/sensitivity),
and the confusion noise from  unresolved galactic (Nelemans \etal 2001) and
extragalactic (Farmer \& Phinney 2003) WD--WD binaries. 
Notice that extreme mass-ratio inspirals (EMRI) could also 
contribute to the confusion noise in the mHz frequency range 
(Barack \& Cutler 2004). 

\section{Results}

\subsection{Coalescence rates}

\begin{figure}
\centerline{\psfig{file=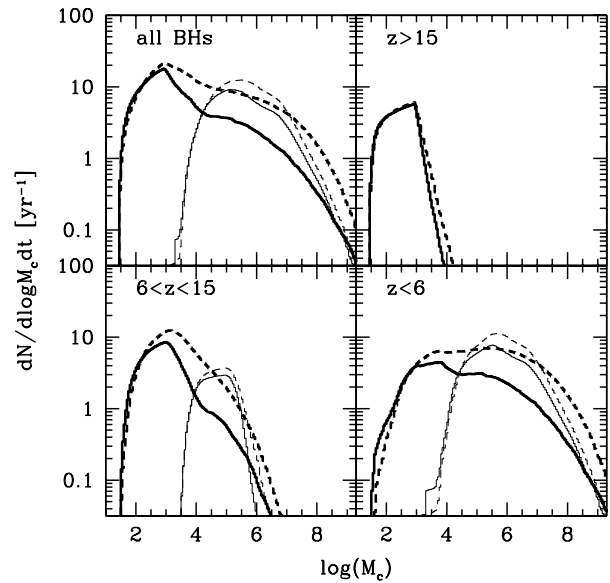,width=84.0mm}}
\caption{Number of MBHB coalescences per observed year at $z=0$, per unit 
log chirp mass, 
in different redshift intervals. {\it Solid lines}: GW recoil 
neglected; {\it dashed lines}: extreme GW recoil included.
{\it Thick lines}: VHM model; {\it thin lines}: BVRlf model.}
\label{mrate}
\end{figure}

Figure \ref{mrate} shows the number of MBH binary coalescences per unit 
${\rm log}{\cal M}$ per unit {\it observed} year, $dN/d{\rm log}{\cal M} dt$, 
predicted by the two models that we have considered, for both cases where
recoil is neglected and 
extreme recoil is taken into account. Each panel shows the rates for 
different redshift intervals. 
Note that when extreme recoil in included, 
the rate predicted by the BVRlf model at any 
redshift is only marginally affected, 
while the VHM model is more sensitive to the GW recoil:
at $z>15$, GW kicks do not affect the coalescence rate; 
on the contrary, at $z<15$, the rate drops by a factor of 
$\sim 3$ for ${\cal M}\gsim 10^3\msun$, 
if extreme kicks are included in the evolution. This is related to the fraction
of seeds that experience multiple coalescences during the MBH assembly
history. We can schematically think of the assembly history as a 
sequence of coalescence rounds, as also recently suggested by Schnittman (2007). After each round 
extreme recoil depletes a large fractions of remnants, and the relative
importance of each subsequent round drops accordingly.
In the VHM model, about 65\% of the 
remnants of the first round will undergo a second round of coalescences,
so the second round has an important relative weight in the computation of the
total rate. When extreme recoil is taken into account, a large fraction of 
the first round remnants is ejected from their hosting halos. We
find that the effective fraction of remnants that can experience
a second coalescence drops to $\sim30\%$. This is the reason why the number
of coalescences involving light black holes (${\cal M}<10^3 \msun$) does not
drop at any redshift, while the number of coalescences involving more
massive binaries drops by a factor $\approx 3$. In the BVRlf scenario
seeds are rarer, and the fraction of first coalescence remnants that
participate to the second round is around 25\%; switching on
the extreme recoil has a significantly smaller 
impact on the global rate in this case.
Moreover, in this model seeds are more massive and the bulk of 
merging events happens at lower redshift, where the hosting halo 
potential wells are deeper and consequently larger kicks are needed
to eject the coalescence remnants. 
As a matter of fact, the seed abundance sets the mean number of major 
mergers that a seed is expected to undergo during the cosmic history, 
and this basically sets the ability of extreme kicks to reduce 
the coalescence rate. 

%%%%%%%%%%%%%%%%%%%%%%%%%%%%%%%%%%%%%%%
%%%%%%%%%%%%%%%%%%%%%%%%%%%%%%%%%%%%%%%

\subsection{LISA detection rate}
%%%%%%%%%%%%%%%%%%%%%%%%%%%%%%%
\begin{figure}
\centerline{\psfig{file=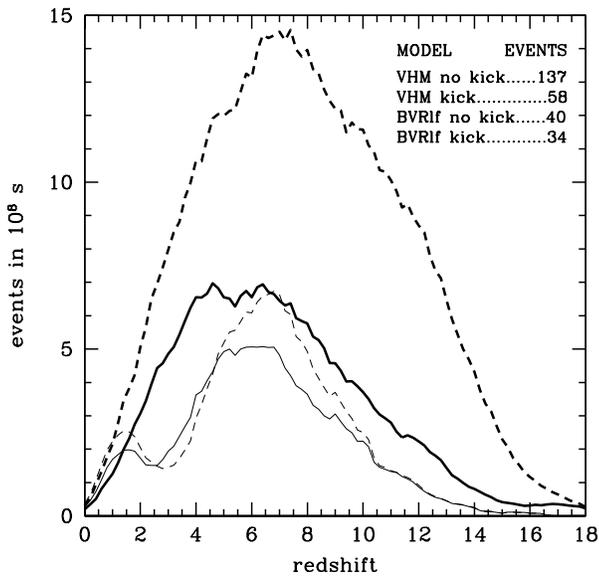,width=84.0mm}}
\caption{Redshift distribution of MBHBs resolved with 
$S/N>5$ by {\it LISA} in a 3-year mission. Line style as in figure \ref{mrate}.
The top-right corner label lists the total number of expected detections.
}
\label{zdist}
\end{figure}
%%%%%%%%%%%%%%%%%%%%%%%%%%%%%%%
%%%%%%%%%%%%%%%%%%%%%%%%%%%%%%%%%%%
\begin{figure}
\centerline{\psfig{file=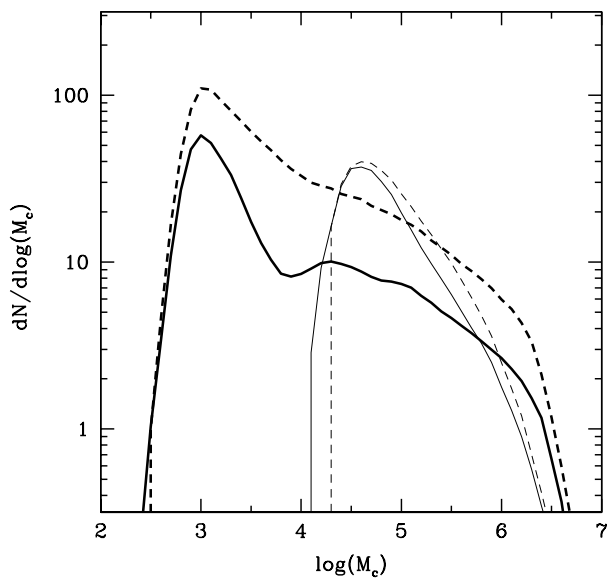,width=84.0mm}}
\caption{Chirp mass function of MBHBs resolved with $S/N>5$ 
by {\it LISA} in a 3-year mission. 
Line style as in figure \ref{mrate}. 
All curves are normalized such as the integral 
in $d\log({\cal M})$ gives the number of detected events.
}
\label{m1dist}
\end{figure}
%%%%%%%%%%%%%%%%%%%%%%%%%%%%%%%%%%%%
We now discuss how the number of GW sources detectable 
by {\it LISA} is influenced by extreme GW recoils. 
To facilitate the comparison with our previous works, 
all the results shown here assume an observation time of 3 years, a 
sharp low-frequency wall at $10^{-4}$ Hz in the instrumental sensitivity 
(see Sesana et al. 2007), and a detection threshold $S/N=5$ (see equation \ref{eqSN}); 
the confusion noise includes only galactic and extragalactic white dwarfs and
ignores a possible contribution from EMRIs (Barack \&
Cutler, 2004). At the end of this section, we will briefly discuss the impact of
the former assumptions on the number of detectable sources. 
Figure~\ref{zdist} shows the redshift distribution of MBHBs detected by {\it LISA}. 
The effect of extreme GW recoils on the source number 
counts drastically depends on the abundance and nature of the seeds,
along the lines discussed in the previous section. In the VHM 
model, the number of detectable sources drops by a factor $\sim 60\%$, and
the number of the potential {\it LISA} detections is reduced from
$\approx 140$, if the recoil is neglected, to $\approx 60$, if extreme 
recoil is included. Vice versa, the detection rate predicted by the BVRlf model
is only weakly affected by the extreme recoil prescription, and it
drops by about $15\%$ (from 40 to 34 events in 3 years of observation). 
Note that though the overall number of coalescences in the VHM
model decreases only by about $25\%$ when extreme recoil is considered,
the number of {\it LISA} detections is reduced by a much larger
factor. This is because if the seeds are light, {\it LISA} can
not detect the bulk of the first coalescences of light
binaries happening at high redshift, that are responsible
for the major contribution to the coalescence rate and
are not affected by the recoil. {\it LISA}
can observe later events, involving more massive binaries,
that are largely suppressed by the MBH depopulation 
due to extreme GW kicks. In the BVRlf
model, on the other hand, seeds are more massive, and the second
coalescence round is less important; in this case, the {\it LISA} sensitivity
is sufficient to observe almost all the first coalescences, and
the number of detections is only mildly reduced. 
As the kicks affect
the merger rate starting from the second round, its signature 
consists in a slight decrease of the mean chirp mass
of the detected binaries, see figure \ref{m1dist}.
  
%%%%%%%%%%%%%%%%%%%%%%%%%%%%%%%%%%%%%%%
\begin{figure}
\centerline{\psfig{file=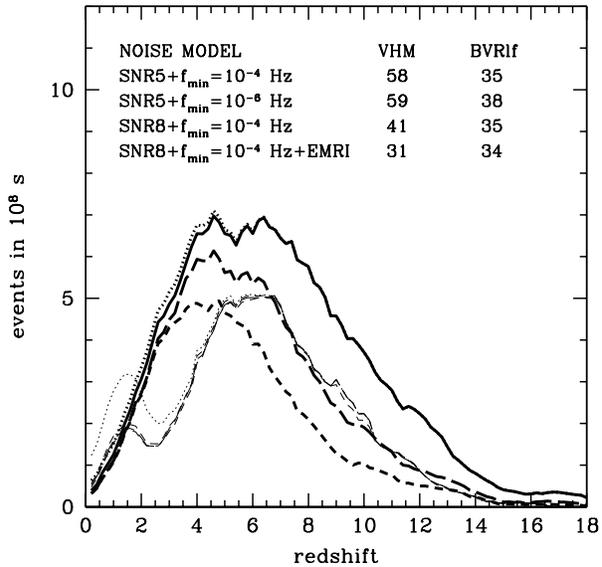,width=84.0mm}}
\caption{Impact of {\it LISA} sensitivity details on the number of detected
sources. {\it Solid lines}: $S/R=5$ and sensitivity cut-off at $f=10^{-4}$ Hz;
{\it long--dashed lines}: $S/R=8$ and sensitivity cut-off at $f=10^{-4}$ Hz; 
{\it short--dashed lines}: $S/R=8$, sensitivity cut-off at $f=10^{-4}$ Hz and
EMRI confusion noise added; {\it dotted lines}: $S/R=5$ and sensitivity cut-off at 
$f=10^{-6}$ Hz. {\it Thick lines} are for VHM model, {\it thin lines}
are for the BVRlf model. The
number of detected MBHBs in a 3 year observation, under the different 
detection assumptions, are also listed. Extreme recoil is assumed.}
\label{lisanoise}
\end{figure}
%%%%%%%%%%%%%%%%%%%%%%%%%%%%%%%%%%%%%%%

We emphasize here two aspects 
(i) at time it is not clear if {\it LISA} will be able to shed light on the importance 
of recoil in MBH assembly, even in this extreme case, since the uncertainty introduced in the 
number counts is at most of a factor of $\sim 3$, comparable with uncertainties
due to our ignorance in the MBH accretion history and in the
detailed dynamics of MBHBs (see, e.g., discussion in Sesana et al. 2007);
(ii) on the other hand, this fact confirm that
MBHBs are {\it LISA} safe targets;
since extreme recoil effects increase with the seed abundance, 
we expect the drop in the detections to be more significant for
those scenarios that predict a larger number of sources. 
In figure \ref{lisanoise} we show how different assumptions on
the detection threshold, the instrumental noise below $10^{-4}$ and the
confusion noise from EMRI affect the LISA detection rates. If seeds are
massive, the results shown in figure \ref{zdist} are hardly affected. 
If seeds are light, EMRI confusion noise and a more conservative detection 
threshold, say $S/N=8$, can halve the number of sources detected by LISA. For 
both scenarios, extending the LISA sensitivity window below $10^{-4}$ has also 
minimal effect on the number of detections

\section{Discussion}

Here we have considered two specific MBH assembly models, representative
of two different MBH seed formation scenarios. However our findings 
can be considered, at least qualitatively, valid in general.
Given the size and the abundance of the seeds, our 'coalescence round' 
picture depends on 
 the details of the models. For example in the VHM model we 
checked that by changing the accretion
prescription (see Volonteri, Salvaterra \& Haardt 2006) 
the total number of events would change by 
a factor of two (note that the accretion 
prescription considered in the models described in the previous section 
gives the minimum number of coalescences); 
however the relative weights of the different coalescence rounds do 
not change significantly. So we can safely conclude 
that a decrease $\gsim 50\%$ in the expected {\it LISA} sources 
should be a general trend for all those models
in which the MBH assembly starts from light seeds at high redshift.
In this class of models the number of predicted coalescing 
events is so high ($\gsim 100$ yr$^{-1}$) that at least a few tens 
of MBHBs should be guaranteed  {\it LISA} sources.
On the other hand, extreme recoils should not be an issue 
at all for {\it LISA} if the MBH seeds are massive and/or rare. 
We remark here that in the BVRlf model 
we assumed MBH seed formation to be efficient 
only in metal free halos with virial temperatures 
$T_{\rm vir} \gsim 10^4$K, i.e, we have considered atomic hydrogen to 
be the only coolant.  Assuming efficient molecular hydrogen gas 
cooling (e.g. Koushiappas, Bullock \& Dekel 2004) the number of seed 
MBHs increases by an order of magnitude (and being the
seeds massive, {\it LISA} would be able to detect the
first coalescence round), and the GW kick would not be an issue at all.
Relying on this results, the estimate of $\sim10$ detections in three years 
predicted by the BVRhf model described in Sesana et
al. 2007 does not change under the assumption of extreme recoils
(seeds are heavy and rare), and can be considered a robust 
 {\it LISA} detection lower limit.
To conclude, in Sesana et al. 2007 we explored different MBH assembly 
scenarios to quantify the imprint of the MBH seed prescription on the
{\it LISA} data stream. Motivated by recent studies on extreme GW recoils, we 
have quantified in this letter their impact on the MBHB coalescence 
and on the {\it LISA} detection rate, confirming that the detection of 
at least $\sim10$ coalescing binaries in a 3 year mission is a 
robust prediction even considering extreme GW recoils. 

%\acknowledgments
\vspace{+0.5cm}
I would like to thank M. Volonteri for having provided the outputs 
of the merger trees used in this study and A. Vecchio and P. Bender 
for helpful discussions and comments on the manuscript. 

{}

\end{document}